\documentclass{aa}

\usepackage{graphicx}
\usepackage{txfonts}
\usepackage{caption}
\usepackage{subcaption}
\usepackage{amsmath,amsfonts,bm}
\usepackage{comment}
\begin{document} 

   \title{The effect of AMR and grid stretching on the magnetized CME model in Icarus}


   \author{T. Baratashvili \inst{1}, S. Poedts \inst{1,2}
          }

   \institute{Department of Mathematics/Centre for mathematical Plasma Astrophysics, 
             KU Leuven, Celestijnenlaan 200B, 3001 Leuven, Belgium. 
             \email{tinatin.baratashvili@kuleuven.be}
             \and
             Institute of Physics, University of Maria Curie-Sk{\l}odowska, 
             ul.\ Radziszewskiego 10, 20-031 Lublin, Poland}

\date{Accepted: December 17, 2023}
\titlerunning{The effect of AMR and grid stretching on spheromak}
\authorrunning{Baratashvili et al.}
 
    \abstract
   {Coronal mass ejections (CMEs) are the main driver of solar wind disturbances near Earth. When directed towards us, the internal magnetic field of the CME can interact with the Earth's magnetic field and cause geomagnetic storms. In order to better predict and avoid damage coming from such events, the optimized heliospheric model Icarus has been implemented. Advanced numerical techniques, such as gradual radial grid stretching and solution adaptive mesh refinement (AMR) are implemented in the model in order to achieve better performance and more reliable results. }
   {The impact of a CME at Earth is greatly affected by its internal magnetic field structure. The aim of this work is to enable modelling the evolution of the magnetic field configuration of the CME throughout its propagation in Icarus. Thus, a magnetized coronal mass ejection model is implemented in Icarus. Such CME model is more realistic than the already available simple hydrodynamics cone CME model and will allow studying the evolution of the magnetised CME during its interactions with the solar wind.The focus of the study is on the global magnetic structure of the CME and its evolution and interaction with the solar wind.} 
   {The magnetized CME model that is implemented in Icarus is the Linear Force-Free Spheromak and is imported from EUHFORIA. The simulations with the spheromak model are performed for different effective resolutions of the computational domain. Advanced techniques, such as grid stretching and AMR are applied. Different AMR levels are applied in order to obtain high resolution locally, where needed. The original uniform middle and high resolution simulation results are also shown, as a reference. The results of all the simulations are compared in detail and the wall-clock times of the simulations are provided.}
   {The results from the performed simulations are analyzed. The co-latitudinal magnetic field component is plotted at 1~AU for both Icarus and EUHFORIA simulations. The time-series at Earth (L1) of the radial velocity, the density and the different magnetic field components are plotted and compared. The arrival time is better approximated by the EUHFORIA simulation, with the CME shock arriving 1.6 and 1.09~hours later than in the AMR level 4 and 5 simulations, respectively. The profile features and variable strengths are best modelled by Icarus simulations with AMR level 4 and 5. The uniform, middle resolution simulation with Icarus took 6.5~hours wall-clock time, while with EUHFORIA the most similar setup takes 18.5~hours, when performed on 1 node with 2 Xeon Gold 6240 CPUs@2.6 GHz (Cascadelake), 18 cores each, on the Genius cluster at KU Leuven. The Icarus simulation with AMR level 4 took only 2.5~hours on the same computer infrastructure, while showing better resolved shocks and magnetic field features, when compared to the observational data and the referene uniform simulation results. }
   {The results from different Icarus simulations in Icarus are presented using results form the EUHFORIA heliospheric modelling tool as a reference. The arrival time is closer to the observed time in the EUHFORIA simulation, but the profiles of the different variables show more features and details in the Icarus simulations. The simulations with AMR levels 4 and 5 showed the most detailed results. Considering the small difference in the modelled results, and the large difference in computational resources, the AMR level 4 simulation is considered to have performed the best. The gradients in the AMR level 4 results are sharper than those in the uniform simulations with both EUHFORIA and Icarus, while the AMR level 4 effective resolution  is the most comparable to the standard resolution runs. The AMR level 3 simulation is 15 and 41 times faster than the Icarus and EUHFORIA uniform simulations, respectively, while the AMR level 4 simulation is $\sim~$3 and 7 times faster than the uniform simulations, respectively. 
   }


   \maketitle
%

\section{Introduction} \label{section:introduction}
Coronal mass ejections (CMEs) are large magnetized plasma clouds erupting from the Sun. Such clouds have masses of $10^{13}\;$g up to $10^{16}\;$g and erupt with velocities ranging from $\sim 100\;\text{km s}^{-1}$ to $\sim 3 000\;\text{km s}^{-1}$ (based on SOHO/LASCO measurements), thus carrying enormous momentum and often causing strong shocks during their evolution in the inner heliosphere. CMEs, when observed by in-situ measurements, are usually referred to as interplanetary CMEs, ICMES. Such an ICME has a complex plasma and magnetic structure, consisting of a leading shock front and a turbulent magnetosheath followed by the magnetic plasma cloud, which often resembles the magnetic field configuration it had during the eruption \citep{Kilpua2017}. The topology of the magnetic cloud, the structure and linkage of its magnetic field, is not fully understood yet and probably varies from case to case. Various studies focus on the radial evolution of the CMEs comparing the magnetic field structure at different phases of the evolution (\cite{al-haddad2019}, \cite{Scolini2022}). ICMEs are the main drivers of the disturbances studied in space weather research. When directed to Earth, the impact of a CME is notable. Such impacts can cause strong geomagnetic storms, induce electric currents in power grids and even disrupt the transmission. During "The Great Qu\'ebec Blackout", Qu\'ebec witnessed the power of a strong CME on March 13, 1989, and they were trying to deal with the consequences for 9 hours.
The damage caused by regular events, especially during phases of maximum activity of the Sun, has been estimated to accumulate to an economic loss of 10 billion\;€/year \citep{press_reference}. The potential loss from the future strong Earth-directed events is increasing continuously, since our dependence on telecommunications, navigation and electronic systems is increasing day-to-day. 
In 2019, the National Threat and Hazard Identification and Risk Assessment (THIRA) of the US Federal Emergency Management Agency (FEMA) identified space weather as one of the two threats that could potentially disturb our society globally, the other one being a pandemic(!) \citep{THIRA2019}.

CMEs occur only a few times a week during solar minima, but during solar maxima, the eruptions are far more frequent, several of them can occur per day \citep{Park2012}. In order to minimize the damage from the arriving ICMEs, realistic and fast space weather forecasting tools are necessary. Currently, a number of such forecasting tools exist. Examples of physics based forecasting tools are ENLIL \citep{Odstrcil2004}, EUHFORIA \citep{Pomoell2018}, SUSANOO-CME \citep{Shiota2016}, and AWSoM \citep{Vanderholst2014}. These tools model the solar wind with superposed CMEs on it, propagating them to the Earth and beyond. Currently, ENLIL and EUHFORIA are used in operational settings, performing daily simulation runs to model the solar wind configuration and the CME evolution. The most simple model for a CME is the so-called cone CME model, which represents a homogeneous hydrodynamic plasma cloud that is injected self-similarly and subsequently evolves in the heliosphere. The advantage of this model is that it is simple and quite efficient. Cone CMEs can model the evolution of CME shocks and predict their arrival time at Earth or other locations in the inner heliosphere. However, this model does not take into consideration the internal magnetic field of the CME. As a consequence, the geo-effectiveness of its impact, coming from its interaction with the magnetic field of the Earth, can not be estimated with this model. In order to predict whether a CME will cause a severe geomagnetic storm upon its arrival at Earth or not, the sign of $B_z$ component of the CME magnetic field, which is the component perpendicular to the equatorial plane, must be modelled accurately. When $B_z$ is negative upon arrival, it is anti-parallel to the magnetic field of the Earth, and such a configuration causes substantial magnetic reconnections and severe geomagnetic storms \citep{Kilpua2017}. Therefore, in order to model the $B_z$ component of the magnetic field of the CME, it is important to incorporate magnetized CME models. The simplest magnetized CME model consists of a spheromak magnetic field configuration. Its shape is similar to the cone CME model, but it has an internal magnetic field. In EUHFORIA, a linear force-free spheromak CME model has been introduced already and it has been used extensively to study different CME events (see e.g., \cite{Verbeke2019}; \cite{Scolini2019}; \cite{Scolini2020b}; \cite{Verbeke2022b}). In the present paper, we consider the same linear force-free spheromak CME model and implement it in the recently introduced heliospheric model Icarus \citep{Verbeke2022}.  

Icarus is a 3D MHD heliospheric wind and CME evolution model implemented in the framework of MPI-AMRVAC (\cite{keppens2003}, \cite{Xia2018}). MPI-AMRVAC is a parallelized architecture, suited for hydro- and magnetohydrodynamics (HD and MHD, respectively) applications. In previous studies, the cone CME was implemented and used for modelling the evolution of CME shocks in the heliosphere. The novelty of Icarus lies in the advanced numerical techniques available in the code that enable better and faster simulations. \cite{Baratashvili2022} examined the effect of radial grid stretching and adaptive mesh refinement (AMR) in the domain on the shock strength of the CMEs and their arrival times, while comparing the wall-clock timings needed for the simulations. Since the interactions of the internal magnetic field of the CME with the magnetic field of the background solar wind play an important role in the CME evolution and its propagation in the heliosphere, the magnetized CME model, namely spheromak was now also implemented in Icarus. 

The aim of the present paper is to quantify and demonstrate the performance of the new magnetized CME model in Icarus. Different AMR strategies are applied to the CME to refine the grid at the varying location of the CME during its propagation, in order to obtain high resolution simulations, and coarsen the grid again after the CME has passed, in order to increase the efficiency of the simulation. For validating the spheromak CME model, the CME that occurred on the 12$^{th}$ of July, 2012 was selected as a typical event to model. This specific event was studied before with the cone CME model in Icarus \citep{Baratashvili2022}, and with both the cone and spheromak CME models in EUHFORIA \citep{Scolini2019}.   

The remainder of the present paper is organized as follows. Section~\ref{section:Icarus_description} describes the numerical model of Icarus. The implementation of the new magnetized CME model is described in detail in Section~\ref{section:spheromak_model}. Then we introduce the simulation set-up and the describe the event in Section~\ref{section:setup}. Sections~\ref{section:results} and \ref{section:discussion_conclusion} are devoted to the presentation and analysis of the simulation results and the discussion and conclusions, respectively. \\

\section{Numerical model} \label{section:Icarus_description}
Icarus is a 3D MHD heliospheric model recently developed at the Center for mathematical Plasma Astrophysics (CmPA), KU Leuven \citep{Verbeke2022}. Icarus is implemented in the framework of MPI-AMRVAC \citep{Xia2018}, which has a heavily parallelized architecture, solving different sets of partial differential equations with various numerical methods and limiters. Different numerical scheme and flux limiter combinations were examined in the Icarus setting to study the effect on the simulation results and the efficiency \citep{Baratashvili2022sungeo}. The experiments were performed for low, middle and high resolution equidistant grids (see Table \ref{tab:resolutions}), in combination with the HLL, HLLC and TVDLF numerical methods and `vanleer' \citep{vanleer1974}, `koren' \citep{Koren1993}, `woodward' \citep{vanleer1977} and `minmod' \citep{Yee1989} flux limiters. \cite{Baratashvili2022sungeo} summarized all 36 simulations and compared the timing and the shock sharpness for each simulation. As a result, the combination of the TVDLF numerical scheme and the `woodward' flux limiter was selected as the most efficient. These settings are fixed as default in all the Icarus simulations presented in this paper. 

The Icarus heliospheric domain is similar to that in EUHFORIA and covers the radial distances from 0.1~AU to 2~AU, 360$^\circ$ in the longitudinal direction and 120$^\circ$ in the latitudinal direction, [-60$^\circ$, 60$^\circ$] in co-latitudes, excluding the poles. The output of the EUHFORIA coronal model is used as the input data for the plasma variables at the inner heliospheric boundary for both EUHFORIA and Icarus. In the Icarus simulations, these values are interpolated on the Icarus grid and ideal MHD equations are solved using the obtained values as boundary conditions. In Icarus, unlike EUHFORIA, the reference frame is co-rotating with the Sun, which leads to a true steady solar wind at the end of the relaxation phase (by definition). The time for the relaxation phase is chosen such that the slow wind stream traverses the whole domain from the inner radial boundary to the outward. The parabolic cleaning method is applied to minimize $\nabla \cdot \mathbf{{B}}$ \citep{Dedner2002}. 

After the steady background wind is obtained, CMEs are superposed on it in the simulations. So far, in Icarus the coronal mass ejections are introduced with a simple HD cone model, which is a non-magnetized plasma cloud with homogeneous interior. Thus, the internal speed, density, temperature and pressure are constant.  The geometry is extensively described by \cite{Scolini2018}. All the mentioned possible shapes are included in Icarus, the same way as in EUHFORIA. The implementation of the cone CME model is described in detail by  \cite{Verbeke2022}. 

The goal of the new heliospheric model Icarus is to perform simulations accurately, yet effectively. This implies achieving the same results as the original configuration of the heliospheric domain, while spending less computational resources and less time. For this purpose, a modified computational grid is used compared to the original EUHFORIA heliospheric simulations. Various advanced techniques are available in the MPI-AMRVAC framework, from which the radial gradual grid stretching and adaptive mesh techniques are addressed in Icarus. 

Since the heliospheric domain covers large distances in the radial direction, an equidistant grid is not the most optimal choice. In spherical coordinates, the equidistant grid cells become deformed ever closer to the outer boundaries, and the longitudinal resolution is affected notably. In order to avoid this issue, gradual radial grid stretching is applied \cite{Xia2018}. With this approach, the aspect ratio of the width and the length of the cells is maintained in the whole domain. The choice for the number of cells in the radial direction is the same as given in \cite{Baratashvili2022} and is fixed to $N=60$. The number of cells in each resolution for the equidistant and stretched grid simulations are given in Table~\ref{tab:resolutions}. For the equidistant grid cases, these can directly be translated to the cell sizes, which are constant throughout the domain. For example, in the middle resolution equidistant grid simulation the size of a cell is $\sim 0.00316(6)$~AU in the radial direction and  $1.875^\circ$ in the longitudinal and latitudinal directions. For the stretched cases, the longitudinal and latitudinal resolutions are also constant, but in the radial direction the sizes of the cells vary along the radial direction and are not fixed to a constant number. 

\begin{table} [hbt!]
\begin{tabular}{|l|l| l|}
\hline
Resolution & \multicolumn{2}{|c|}{\# cells[$r,\theta,\phi$]}
\\\hline\hline
 & Equidistant & Radially stretched \\\hline
Low& [300,\;32,\;96]& [60,\;32,\;96]\\\hline
Middle& [600,\;64,\;192]& [120,\;64,\;192]\\\hline
High& [1200,\;128,\;384]& [240,\;128,\;384]\\\hline
\end{tabular}
\caption{List of equidistant (uniform) and radially stretched grid resolutions referred to throughout this paper. }
\label{tab:resolutions}
\end{table}

The second advanced technique exploited in Icarus is called adaptive mesh refinement. The details of block-adaptive AMR in MPI-AMRVAC can be found in \cite{keppens2003}. The AMR can be initiated in the domain by prescribing a criterion for the refinement. Whenever the criterion is met in the simulation, the block is refined to the intended refinement level. Different AMR criteria, suited for heliospheric simulations, are discussed in \cite{Baratashvili2022}. The criterion can be aimed at the CME interior, CME and CIR shocks in the solar wind, the combination of the two, or any other regions of interest in the domain. For example, if the CME tracing refinement criterion is prescribed, the CME will be traced along its propagation from the inner heliospheric boundary to the outer one. The whole block containing the portion of the CME is refined, but once the CME passes the block, it is coarsened to the base resolution of the domain, in order to avoid unnecessary high resolution computational grid regions. The base resolution of the simulation is always fixed to the low resolution given in Table~\ref{tab:resolutions}. 

Since the radial resolution of the stretched grid varies, Table~\ref{table:amr_resolutions_at_l1} summarizes the radial size of the cells at the first Lagrange point L1 with different AMR levels. 

\begin{table}[htb!]
\caption{Resolutions at L1 for different simulations with the stretched grid. No AMR corresponds to the low-resolution simulation.}   
\label{table:amr_resolutions_at_l1}   
\centering            
\begin{tabular}{c c c c c}         
\hline\hline  
No AMR & AMR 2 & AMR 3 & AMR 4 & AMR 5 \\ 
\hline            
    10.7 R$_\odot$ & 5.36R$_\odot$ & 2.68 R$_\odot$ & 1.344R$_\odot$  & 0.672 R$_\odot$ \\
\hline                                  
\end{tabular}
\end{table}

The cell size in every AMR level \textit{n} is $2^{(n-1)}$ times smaller than in the base resolution of the simulations in every spatial direction, thus, in the next AMR level, every refined grid cell (satisfying the criteria) is cut in 8 smaller cells. For example, for a radially stretched grid, the radial cell size at L1 in the AMR level 5 simulation is 0.672$\;R_\odot$, which is 16 times smaller (in every direction) than in the low resolution simulation, corresponding to 10.7$\;R_\odot$ (at L1). Examples for the other AMR levels are given in Table~2.

\section{Magnetized CME model} \label{section:spheromak_model}
The magnetized CME model implemented in Icarus is a Linear Force-Free (LFF) Spheromak model. This is achieved by linking Icarus to the existing spheromak CME model in EUHFORIA \citep{Verbeke2019}. Since the magnetized CME is imported from EUHFORIA and not implemented anew, we will discuss the details of the model only briefly. An extensive discussion of the model and its implementation in EUHFORIA is given in \cite{Verbeke2019}, and the spheromak CME model is also described and discussed by \cite{Kataoka2009} and \cite{Shiota2016}.
\begin{table}[htb!]
\caption{The input parameters for the LFF Spheromak model.}   
\label{table:spheromak_input_pars}   
\centering            
\begin{tabular}{c c c}         
\hline\hline  
Variable & Description & Value range  \\ 
\hline            
    t$_{CME}$ & time of the CME at 0.1~AU & in UT format  \\
    $\theta_{CME}$ & latitude of the center of the CME & [-60$^\circ$, 60$^\circ$] \\
    $\phi_{CME}$ & longitude of the center of the CME & [0$^\circ$, 360$^\circ$] \\
    r$_{CME}$ & radius of the CME & [0-0.1]AU \\
    v$_{CME}$ & speed of the CME & >0 m s$^{-1}$ \\
    $\rho_{CME}$ & density of the CME & >0 kg m$^{-3}$ \\
    T$_{CME}$ & temperature of the CME & >0 K \\
    $\tau_{CME}$ & tilt angle of the spheromak & [0$^\circ$, 360$^\circ$] \\
    H$_{CME}$ & helicity of the CME & +1, -1 \\
    F$_{CME}$ & total toroidal flux of the CME & $\mathbb{R}$\\
\hline                                
\end{tabular}
\end{table}

The model is in fact a modification of the Gibson and Low magnetized CME model \citep{Gibson1998}, which is not force-free. In this force-free version, offered by \cite{Shiota2016}, the CME evolves completely through the inner boundary so that its foot points do not stay attached to the inner heliospheric boundary. Moreover, this LFF Spheromak model implementation pushes it through the inner heliospheric boundary in such a way that positive pressure is guaranteed everywhere, unlike the original Gibson and Low model. The CME is represented by a sphere with a fixed radius when evolving through the inner boundary, thus it is not evolving self-similarly (with growing radius) like the cone model. However, similar to the cone CME model, the plasma variables in the interior are considered to be homogeneous. In order to inject the CME in the heliosphere as a time-dependent boundary condition, each point of the inner heliospheric boundary is checked at every time step in the simulation. When the coordinates of the centre of the CME are given in Cartesian coordinates by (x$_{CME}$, z$_{CME}$, z$_{CME}$), and the coordinates for an arbitrary point at the inner heliospheric boundary are given by (x$_{IHB}$, z$_{IHB}$, z$_{IHB}$), the following condition is checked
\begin{align}
  (x_{CME} - x_{IHB})^2 + (y_{CME} - y_{IHB})^2 + (z_{CME} - z_{IHB})^2 \leq r_{CME}^2,
\end{align}
where r$_{CME}$ is the (fixed) radius of the (moving) CME. If this inequality holds for the given point at a certain time, then it belongs to the CME, if not, then it belongs to the background solar wind and should not be adjusted. 

The magnetic field is defined in the local spherical coordinate system ($r '$, $\theta '$, $\phi '$), where the origin is considered the center of the CME, and it exhibits symmetry in the azimuthal direction ($\phi '$). The magnetic field vector can be expressed as
\begin{align} \label{B_field_vector}
\mathbf{B} = \frac{1}{r' sin\theta'} \Big [\frac{1}{r'} \frac{\partial \psi}{\partial \theta'} \mathbf{\hat{r}'} - \frac{\partial \psi}{\partial r'} \boldsymbol{\hat{\theta}'} + Q \boldsymbol{\hat{\phi}'} \Big ],
\end{align} 
where $\psi$ and $Q$ are scalar potentials that only depend on $r'$ and $\theta'$ \citep{Chandrasekhar1956}, and $\mathbf{\hat{r}'}$, $\boldsymbol{\hat{\theta}'}$, and $\boldsymbol{\hat{\phi}'}$ denote unit vectors in each of the three spatial coordinate directions. The magnetic field is divergence-free by the given definition, and the solution is determined so that $\mathbf{J} \, \times \, \mathbf{B} = 0$, thus force-free. From the azimuthal force balance, the toroidal field must be the function of a poloidal potential, therefore we get $Q = Q(\psi)$. In order to obtain the linear force-free model, we assume $Q(\psi) =  \alpha H \psi$, where $H$ is the constant describing the helicity of the spheromak, and $\alpha$ is the length scale constant, which is determined from the condition that the magnetic field at the surface is zero. After solving the force-free equation, the poloidal potential is obtained as
\begin{align} \label{poloidal_potential}
\psi = \frac{B_0}{\alpha} r' j_1(\alpha r') sin^2\theta',
\end{align}
where $B_0$ is the magnetic field strength and $j_1(x)$ is the spherical Bessel function of order one.  If we expand the magnetic field into the components and use the expression for the poloidal potential, we obtain the magnetic field configuration of the LFF spheromak model
\begin{align}
  B_r' &= 2B_0 \frac{j_1(\alpha r')}{\alpha r'} \cos \theta', \\
  B_\theta' &=-B_0 \Big[  \frac{j_1(\alpha r')}{\alpha r'} + j_1(\alpha r') \Big] \sin\theta', \\
  B_\phi' &= H \cdot B_0 j_1(\alpha r')\sin\theta'.
\end{align}
The magnetic field is assumed to be zero at the spheromak boundary. If $\psi(r=r_{CME}) = 0$, we obtain the condition for the spherical Bessel function of order one
\begin{align*}
j_1(\alpha r_{CME}) = 0
\end{align*}
using the first zero of spherical Bessel function of order one, we obtain $\alpha r_{CME} \sim 4.4934094579$.

The input parameters for the LFF spheromak model also include the parameters characterizing the internal magnetic field configuration, together with the parameters describing the geometry and hydrodynamics characteristics. The overview for the input parameters is given in Table~\ref{table:spheromak_input_pars}, which closely follows the one described by \cite{Verbeke2019}.

\section{Simulation set-up} \label{section:setup}
\subsection{Event description}
In order to validate and examine the implemented magnetized LFF spheromak model, we selected a well-studied CME event of which the impact at Earth was observed clearly. The CME originated from the active region, NOAA AR11520. It followed a strong X1.4 class fare, which was observed by LASCO C2 coronagraph at 16:48UT July 12, 2012. The parameters for modelling this CME with the spheromak model are taken from \cite{Scolini2019} and are summarized in Table~\ref{table:event_parameters}.

\begin{table}[htb!]
\caption{The parameters for the CME event on July 12, 2012.}   
\label{table:event_parameters}   
\centering            
\begin{tabular}{c c}         
\hline\hline  
Variable & Input value \\ 
\hline            
    t$_{CME}$ & 2012-07-12T19:24   \\
    $\theta_{CME}$ & -4$^\circ$  \\
    $\phi_{CME}$ & -8$^\circ$  \\
    r$_{CME}$ & 16.8 R$_\odot$ \\
    v$_{CME}$ & 763 km s$^{-1}$ \\
    $\rho_{CME}$ & 10$^{18}$ kg m$^{-3}$ \\
    T$_{CME}$ & 0.8 x $^6$ K \\
    $\tau_{CME}$ & -135$^\circ$ \\
    H$_{CME}$ & +1 \\
    F$_{CME}$ & 10$^{14}$ Wb\\
\hline                                
\end{tabular}
\end{table}

\subsection{Simulations overview}
As mentioned before, the magnetized CME model is imported directly from EUHFORIA, it is exactly the same and any differences in the simulations are entirely due to the differences in the numerical methods, including the grid stretching and the mesh adaptivity. Therefore, a link has been set up between Icarus and EUHFORIA to facilitate communication between the models whenever the magnetic field components need to be computed. The simulations are performed with a predefined AMR criterion suited for resolving the complex CME interior \citep{Baratashvili2022}.

\subsubsection{Linking Icarus to EUHFORIA}
\begin{figure}[hbt!]
\centering
    \includegraphics[width=0.49\textwidth]{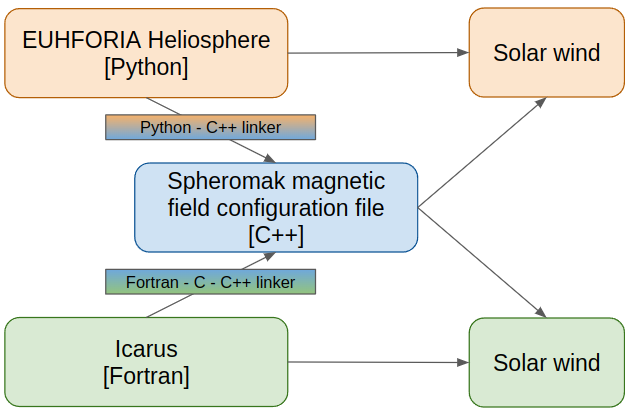}
  \caption{Diagram shows the procedure of importing the spheromak file in Icarus used in EUHFORIA.}\label{fig:diagram_linking}
\end{figure}

The heliosphere model of EUHFORIA is implemented in Python, while the Spheromak CME model is implemented in C++. The heliosphere model is connected to the spheromak model via a Python-C++ linker, in order to compute magnetic field components at each point belonging to the CME. In order to avoid unnecessary duplication and to ensure exactly the same implementation, Icarus is also connected to the same implemented spheromak model in C++. The visual representation of the connections is plotted in Figure~\ref{fig:diagram_linking}.

Icarus is implemented in a Fortran environment. Hence, a different link has been set up to the spheromak code. First, in EUHFORIA, additional static libraries have been created from the existing spheromak magnetic field computation code, intended for linking to Icarus. In Icarus, on the other hand, a Fortran-C++ linker has been added to the compilation settings and the code has been recompiled. The path to the newly generated static library is indicated for linking to the required library. As a result, the compiled code with these modifications links the spheromak computation code to Icarus. Upon injecting the CME at the inner heliospheric boundary in the simulation, the linked library is called from Icarus, which first computes and then returns the magnetic field components at the requested points in the computational domain. The parameters characterizing the CME are set locally, as they do not depend on time. Of course, while passing the location of the intended CME point, the difference in the reference frames of the EUHFORIA and Icarus simulations are taken into account.

\begin{figure*}[hbt!]
     \centering
     \begin{subfigure}[b]{0.33\textwidth}
         \centering
         \includegraphics[width=\textwidth]{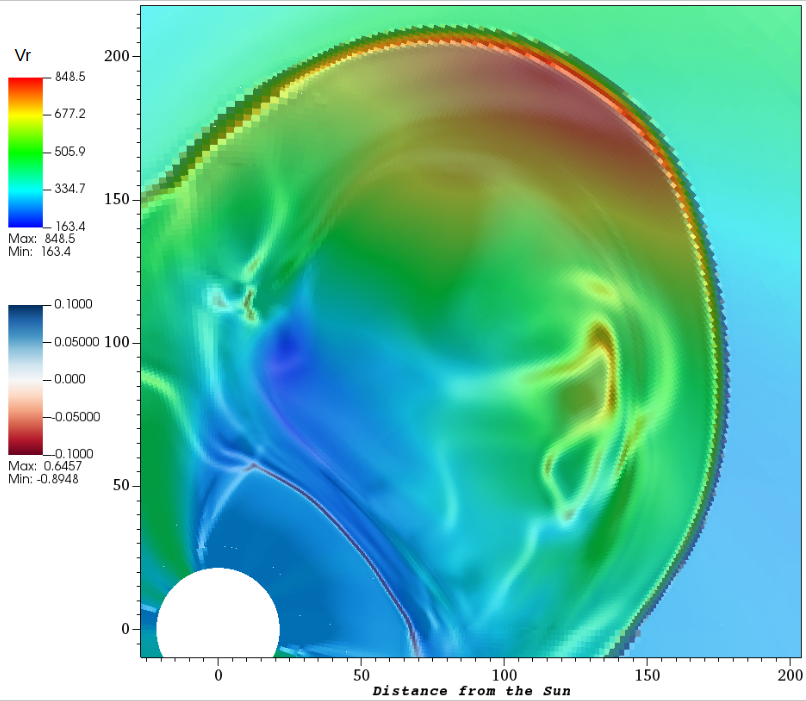}
         \caption{Radial velocity and Div($\mathbf{V}$)}
         \label{fig:amr_vr_div}
     \end{subfigure}
     \hfill
      \begin{subfigure}[b]{0.33\textwidth}
         \centering
         \includegraphics[width=\textwidth]{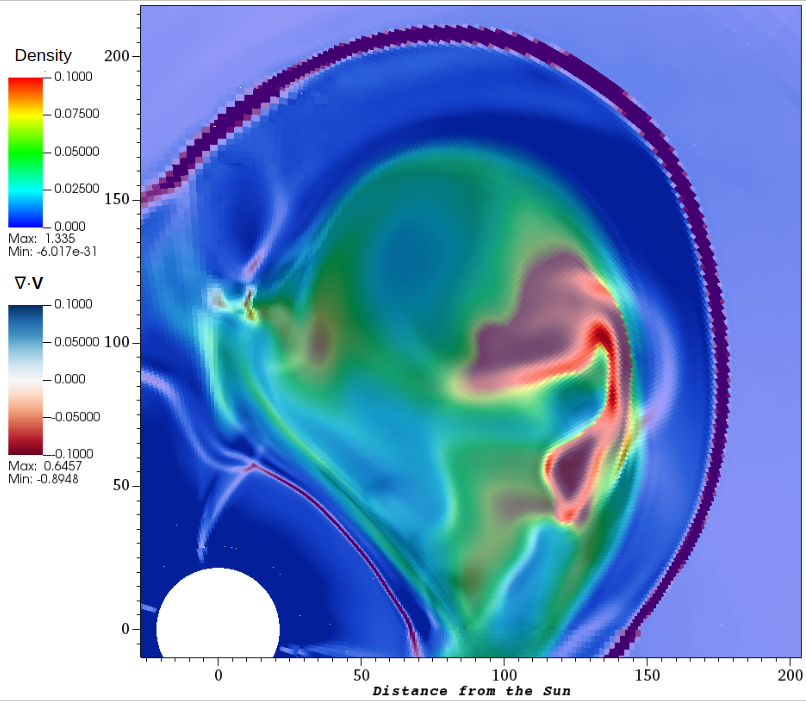}
         \caption{CME density and Div($\mathbf{V}$)}
         \label{fig:amr_trp_div}
     \end{subfigure}
     \hfill
     \begin{subfigure}[b]{0.33\textwidth}
         \centering
         \includegraphics[width=\textwidth]{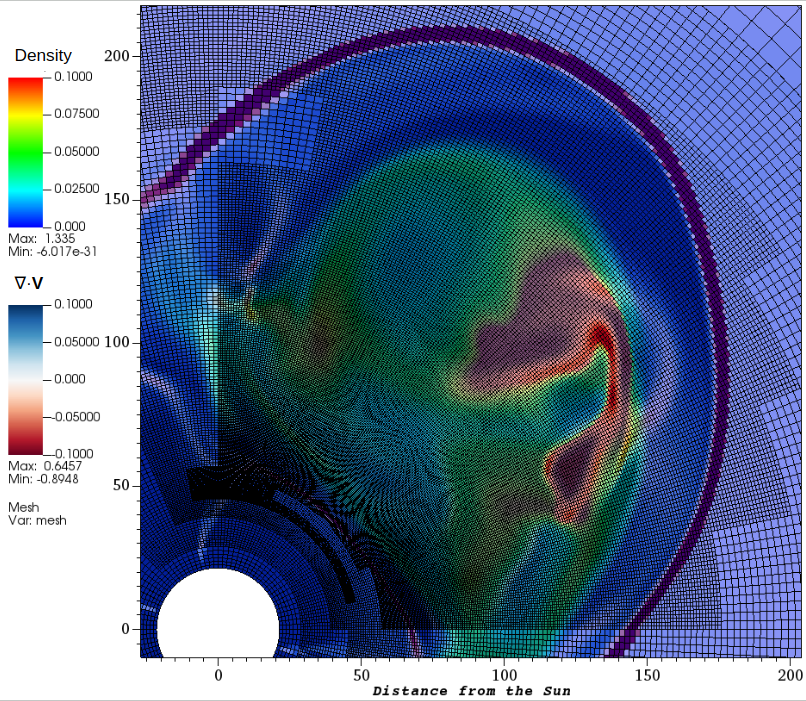}
         \caption{CME density and Div($\mathbf{V}$) with mesh}
         \label{fig:amr_mesh}
     \end{subfigure}
        \caption{Results in the ecliptic plane for the AMR level 4 simulation with the combined AMR criterion. The left panel shows the radial velocity values overlaid with the divergence of $\mathbf{V}$. The middle panel shows the CME density values overlaid with divergence of $\mathbf{V}$. The right panel shows the mesh superposed on the middle figure. }
        \label{fig:amr_criterion_in_ecliptic}
\end{figure*}

\subsubsection{AMR criterion} \label{advanced_techniques}

In order to demonstrate the potential of AMR for resolving the inner magnetic structure of the simulated CMEs, the spheromak CME is injected in Icarus with different resolution grids. The `standard' (for operational use) middle resolution uniform grid is taken as a reference. In order to see if we can improve the results with advanced techniques, the radial grid stretching is combined with AMR. \cite{Baratashvili2022} examines different AMR strategies. The first criteria focuses on the CME interior and is implemented by tracing the density in the heliosphere as the CME propagates towards the outer boundary. The other criterion considered in the paper resolves shocks in the domain by tracing the sign of the divergence of $\mathbf{V}$, hinting at the regions of compression or expansion. The last AMR strategy combines the first two criteria, thus, focuses on the CME interior and its shock-front along its propagation and evolution in the inner heliosphere. The AMR criterion used in all the simulations in this paper is a combination of two criteria, called the `combined' criterion by \cite{Baratashvili2022}. This combined criterion is well suited for studies of the evolution of CMEs as it fully resolves the complex CME models. It is therefore also well suited for the considered test case, since the complex internal magnetic field of the CME is modelled in these simulations. 

Apart from the CME interior, the CME shock is also maximally resolved to accurately model the shock strength upon arrival at L1. The mesh refinement is limited to a narrow bandwidth around the Earth-Sun line for the space weather forecasting purposes. The modelled time series of solar wind parameters at Earth are compared to each other (for different resolutions) and to observational data. This approach further optimizes the simulations and saves a lot of computational resources. A visual representation of how this combined AMR criterion is applied in the domain is given in Figure~\ref{fig:amr_criterion_in_ecliptic}. The figure shows cuts of the 3D simulation in the ecliptic plane. A small portion of the ecliptic plane is zoomed in, including the CME, in order to show more details of the CME coverage. Figure~\ref{fig:amr_vr_div} shows the radial velocity values overlaid with the values of the divergence of $\mathbf{V}$ (used as AMR criterion) with corresponding colour maps. Thus, in this figure the shock-front is distinguished with a red colour, corresponding to negative div($\mathbf{V}$) values, and it is followed by the CME magnetic cloud, depicted with $V_r$ values. 

Figure~\ref{fig:amr_trp_div} shows the CME density profile overlaid with the divergence of $\mathbf{V}$ values with corresponding colour maps. Since the AMR criterion takes into account the CME density and the shock, the combined profile was plotted. The CME density is an independent variable, that only describes the plasma density in the CME interior, and is zero everywhere else in the domain. Hence, the magnetosheath of the CME can be distinguished very easily between the CME density profile and the leading shock. 

Figure~\ref{fig:amr_mesh} shows the same variables as in Figure~\ref{fig:amr_trp_div}, but with the overlaid mesh. The `combined' refinement criterion has the additional restriction of the bandwidth around the Sun-Earth line, which implies that at every time-step the AMR is applied along the longitude where the Earth lies with a given margin $\Delta \phi =\pm 30^\circ$. For example, if Earth is at longitude $\phi = 60^\circ$,  the areas meeting the implemented AMR criterion are refined to the indicated refinement level between longitudes $30^\circ < \phi < 90^\circ$. As a result, the whole area including the CME shock and the CME interior in the heliosphere is not refined to the highest refinement level, but only the regions along the Sun-Earth line.

\section{Results and Discussion} \label{section:results}
The event study was performed in order to validate the integration of the magnetized LFF Spheromak CME model in Icarus. As an input for the EUHFORIA coronal model, we selected the standard synoptic magnetogram from the Global Oscillation Network Group (GONG) on July 12, 2012 at 11:54UT. Then, the output of the coronal model, that is MHD parameters at 0.1~AU,  was used as the inner boundary conditions for the onset of the heliospheric simulations in Icarus, which first extended them radially to 2~AU and then performed an MHD relaxation period of 14 days to obtain a steady background wind. The parameters used for the CME injection from the inner heliospheric boundary are given in Table~\ref{table:event_parameters}.

\begin{figure*}[hbt!]
     \centering
     \begin{subfigure}[b]{0.49\textwidth}
         \centering
         \includegraphics[width=\textwidth]{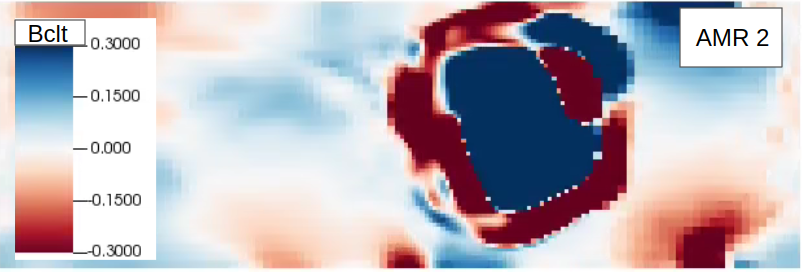}
         \caption{AMR 2}
         \label{fig:bclt_amr2}
     \end{subfigure}
     \hfill
     \begin{subfigure}[b]{0.49\textwidth}
         \centering
         \includegraphics[width=\textwidth]{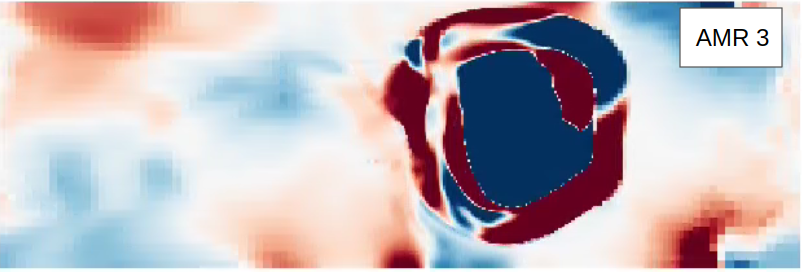}
         \caption{AMR 3}
         \label{fig:bclt_amr3}
     \end{subfigure}
     \hfill
     \begin{subfigure}[b]{0.49\textwidth}
         \centering
         \includegraphics[width=\textwidth]{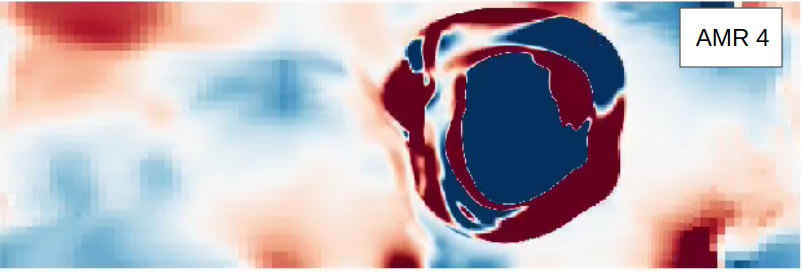}
         \caption{AMR 4}
         \label{fig:bclt_amr4}
     \end{subfigure}
     \hfill
     \begin{subfigure}[b]{0.49\textwidth}
         \centering
         \includegraphics[width=\textwidth]{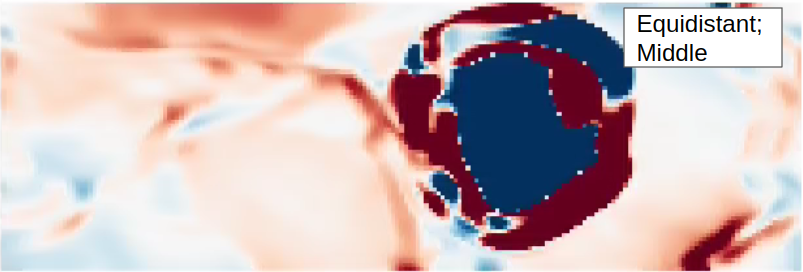}
         \caption{uniform grid; middle resolution}
         \label{fig:bclt_equidistant}
     \end{subfigure}
        \caption{Figures are given at the 1~AU slice. The $B_\text{clt}$ component of the magnetic field is plotted with the same scaling for all the figures for comparison purposes. AMR levels 2, 3 and 4 are used in the first three figures. The last figure shows results for the uniform middle resolution grid simulation. }
        \label{fig:Bclt_at_1au}
\end{figure*}

Several simulations were performed on different computational grids to compare the performance of the advanced techniques with that of the original uniform grid simulations. The co-latitudinal component of the magnetic field obtained in the simulations is plotted in Figure~\ref{fig:Bclt_at_1au}. The $B_\text{clt}$ values are saturated with $(-0.3, 0.3)$ nT, in order to emphasize the difference between regions with negative and positive values. For all the simulations with the stretched grid and AMR, the AMR criterion  is fixed to the combined refinement criterion described in Section~\ref{advanced_techniques}. Figure~\ref{fig:bclt_amr2} represents the simulation result obtained with AMR level 2, with one refinement level. Figure~\ref{fig:bclt_amr3} represents the results obtained with AMR level 3. Figure~\ref{fig:bclt_amr4} corresponds to the simulation with AMR level 4, and Figure~\ref{fig:bclt_equidistant} shows the results obtained without radial grid stretching, nor AMR. In the latter simulation, the original uniform middle resolution grid is used. For each simulation, the same snapshot is taken, which is chosen in the time interval in which the spheromak is crossing the sphere with radius of 1~AU.

The interior structure of the spheromak can be well distinguished from the background solar wind at 1~AU. The background wind is not very well resolved in the simulations with the stretched grid and AMR. As a result, more details and smaller-scale structures can be distinguished in the background wind simulated in the uniform grid. The reason for this is that the prescribed AMR criterion does not take into consideration the background solar wind, as it focusses on the CME itself. Therefore, the background wind is resolved to the base resolution of the heliospheric domain, which corresponds to the \textit{No AMR} resolution from Table~\ref{table:amr_resolutions_at_l1}. Remark that the cell sizes are rather large here, which leads to such smooth profiles. Since the resolution is uniform everywhere in the uniform grid simulation, the solar wind is better resolved at 1~AU in that case. On the contrary, when comparing the spheromak CME interior in the different AMR simulations, more detailed structures are resolved with more AMR levels. The boundaries between opposite polarity regions are larger in the AMR level 2 run than in the AMR level 4 run. The shapes and structures lack details in the AMR level 2 case compared to the AMR level 3 and 4 cases. The overall structure of the magnetized CME is similar in all the simulations, as expected. When increasing the number of AMR levels, a higher resolution is obtained locally as the resolution is doubled for each higher level, which leads to better resolved gradients and local variations between the opposite polarity regions. This is especially important to study the evolution of the magnetized CMEs in the heliosphere, determined by its interactions with the also magnetized background solar wind, and in particular, the changes in the complex internal magnetic field structure. The uniform grid simulation result is mostly comparable to the AMR level 3 result, and the features are finer and better resolved in AMR level 4 simulation result. This, in fact, is expected, since the longitudinal and latitudinal resolutions are not affected by radial grid stretching. In AMR level 3, the resolution is one level higher than middle uniform simulation in the latitude-longitude plane, but considering the lower radial resolution, AMR level 3 shows slightly smoother profiles at 1~AU. 

\begin{figure}[hbt!]
\centering
    \includegraphics[width=0.49\textwidth]{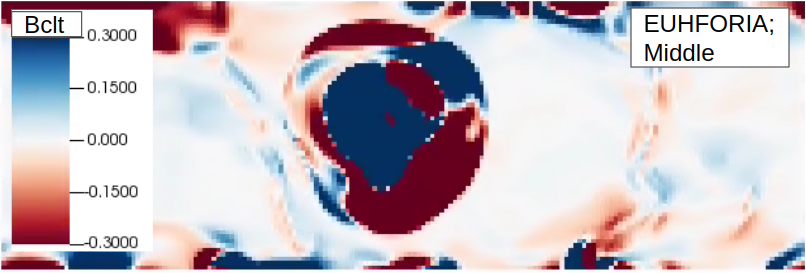}
  \caption{$B_\textbf{clt}$ in the EUHFORIA simulation, also at the 1~AU slice. The magnetic field is scaled similar to the results for the Icarus simulations shown in Figure~\ref{fig:Bclt_at_1au}. }\label{fig:Bclt_at_1au_euhforia}
\end{figure}

Figure \ref{fig:Bclt_at_1au_euhforia} shows the corresponding result for $B_\text{clt}$ component at 1~AU in the EUHFORIA simulation with the middle resolution uniform grid. A snapshot corresponding to the most similar time as for the shown Icarus results was chosen from the data set. The small differences in the CME interior is caused by the fact that the CME does not arrive exactly at the same time at L1 in Icarus and EUHFORIA, therefore a slightly different cut through the 3D spheromak structure is captured at the given time in the simulation in EUFHORIA.  Notice that the background solar wind magnetic field is different from that in the Icarus results. There are more profiles present near the boundaries in the latitudinal direction. The features are well resolved in the background wind, similar to the Icarus uniform middle resolution grid simulations. Also, the spheromak interior looks similar to that obtained with Icarus results, in terms of negative and positive portions and their alignments. The details and small structures are different in the interior, as expected, since very different numerical methods are used. Details about the EUHFORIA simulation setup with a spheromak CME can be found in \cite{Pomoell2018} and \cite{Verbeke2019}. The fact that the CME locations are not at the same place in Figures~\ref{fig:Bclt_at_1au} and \ref{fig:Bclt_at_1au_euhforia}, is caused by the difference in reference frames (HEEQ in EUHFORIA while corotating in Icarus). The CME is injected at the same location, but since we are plotting a given snapshot in both simulations the whole domain is shifted, including the spheromak. 

The AMR condition was limited for optimizing forecasting scenarios in these simulations. Thus, the refined area in the longitudinal direction was restricted to $\pm 30^\circ$ from the Sun-Earth line, as explained in Section~\ref{advanced_techniques}. Usually, the restriction is also applied to the co-latitudinal direction, $\pm 20^\circ$ from the equatorial plane, to further optimize forecasting simulations. In this case, however, the entire CME was resolved to higher refinement levels with the purpose to show the refined structure better. 


\begin{figure}[hbt!]
    \includegraphics[width=\linewidth]{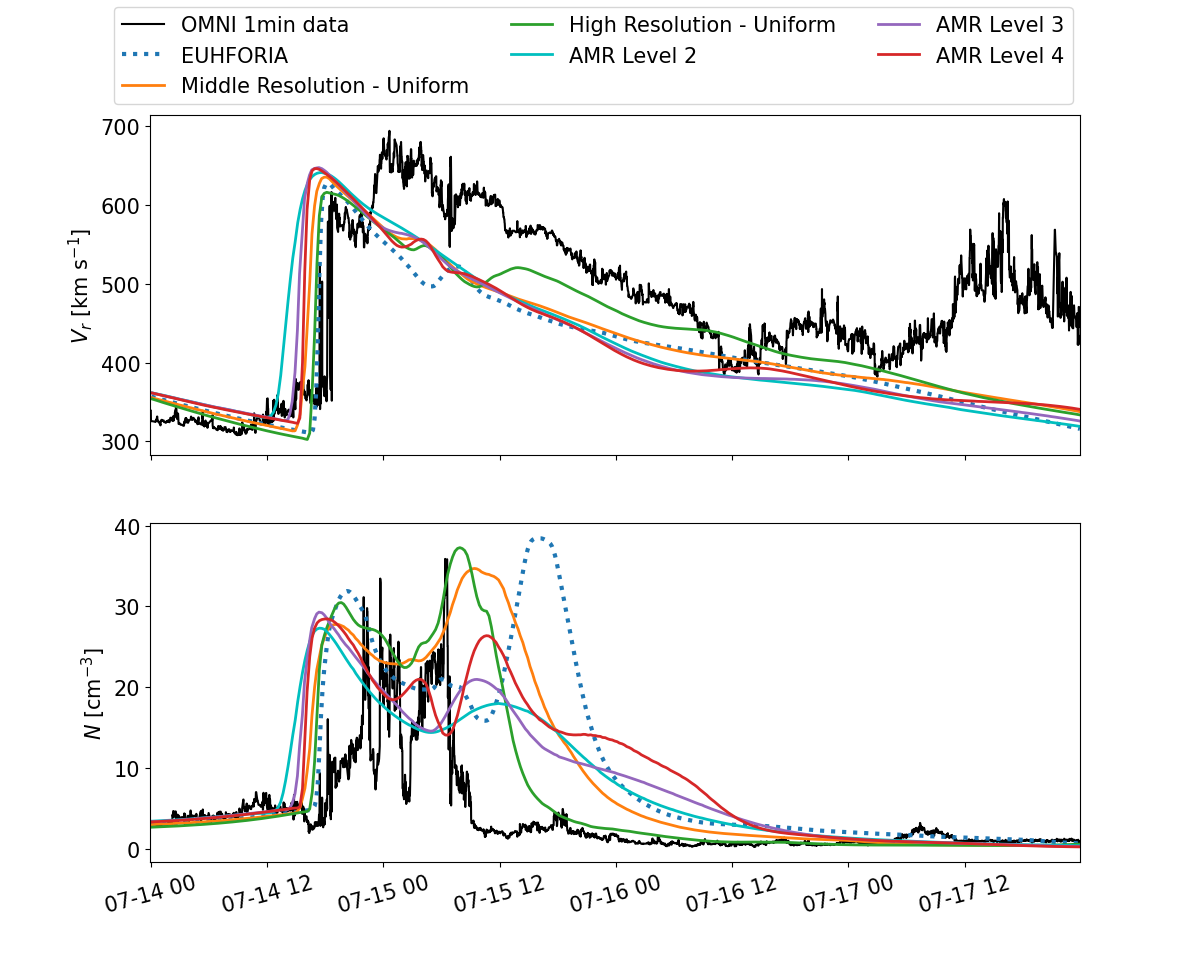}
    \caption{Time series data at Earth. Time profiles for the radial speed and number density values are given for the uniform grid simulations in Icarus and EUHFORIA, together with Icarus simulations using AMR levels 2, 3, and 4.} \label{fig:speed_numberdensity_timeseries}
\end{figure}

Figure~\ref{fig:speed_numberdensity_timeseries} shows the evolution of the radial velocity component (upper panel) and the number density (lower panel) at Earth. The black curve represents the corresponding OMNI 1~min data.  The simulation was performed for a longer time interval than shown, but the dates were trimmed for better visibility, including only a short period before the arrival of the CME and a longer period until the CME has completely passed Earth. The corresponding EUHFORIA middle uniform grid simulation result is plotted with a blue dotted line, and the Icarus middle and high resolution uniform grid simulations with an orange and green solid lines. The Icarus AMR level 2, 3 and 4 simulation results are plotted with cyan , purple and red lines, respectively. The jump in the radial velocity at around noon on July 14 corresponds to the shock arrival at Earth. The shock arrival seems to occur at different times for different simulations with Icarus, at least when the increase of the radial velocity is used as criterion. This is because the shock is more diffuse in the low resolution simulations than in the higher resolution simulations. As a consequence, it occupies the same amount or more cells in the radial direction (e.g.\ 3 to 4 cells for AMR level 2, while mostly 3 cells in AMR level 3), and moreover, the grid cells become smaller by a factor of 2 for each higher AMR level. As a result, when reading out the temporal evolution of the variables near Earth, the CME shock effectively reaches 1~AU earlier than in higher resolution simulations. This is merely because the shock is more diffuse, while in higher resolution simulations it is more localized because it is captured with less grid cells and, moreover, these cells are smaller. Despite this diffusion effect, the shock strength is more or less independent of the resolution and the peak is reached at the same time in the different AMR level simulations. The shock profile is different in middle and high resolution uniform grid simulations, since there, the grid resolution is different in the entire computational domain, which affects the solar wind too as can be seen in the figure before the arrival of the shock. The CME shock arrives slightly later in the EUHFORIA simulation than in the Icarus simulations. This can be explained by the difference of the numerical methods used in the two different heliospheric models. The radial velocity values increase from $\sim300\;$km s$^{-1}$ to $\sim 650\;$km s$^{-1}$. The number density also jumps from $\sim 5\;$cm$^{-3}$ to $\sim 30\;$cm$^{-3}$. Notice also that in the different simulations, the steepness of the curve is different, due to the different radial resolutions at 1~AU. The most steep profile is obtained with the AMR level 4 simulation, followed by the Icarus and EUHFORIA middle uniform grid simulations. As expected, the smoothest profile is modelled by the AMR level 2 simulation, because the radial resolution is the lowest in this case.\\

\begin{figure*}[hbt!]
\centering
    \includegraphics[width=\textwidth]{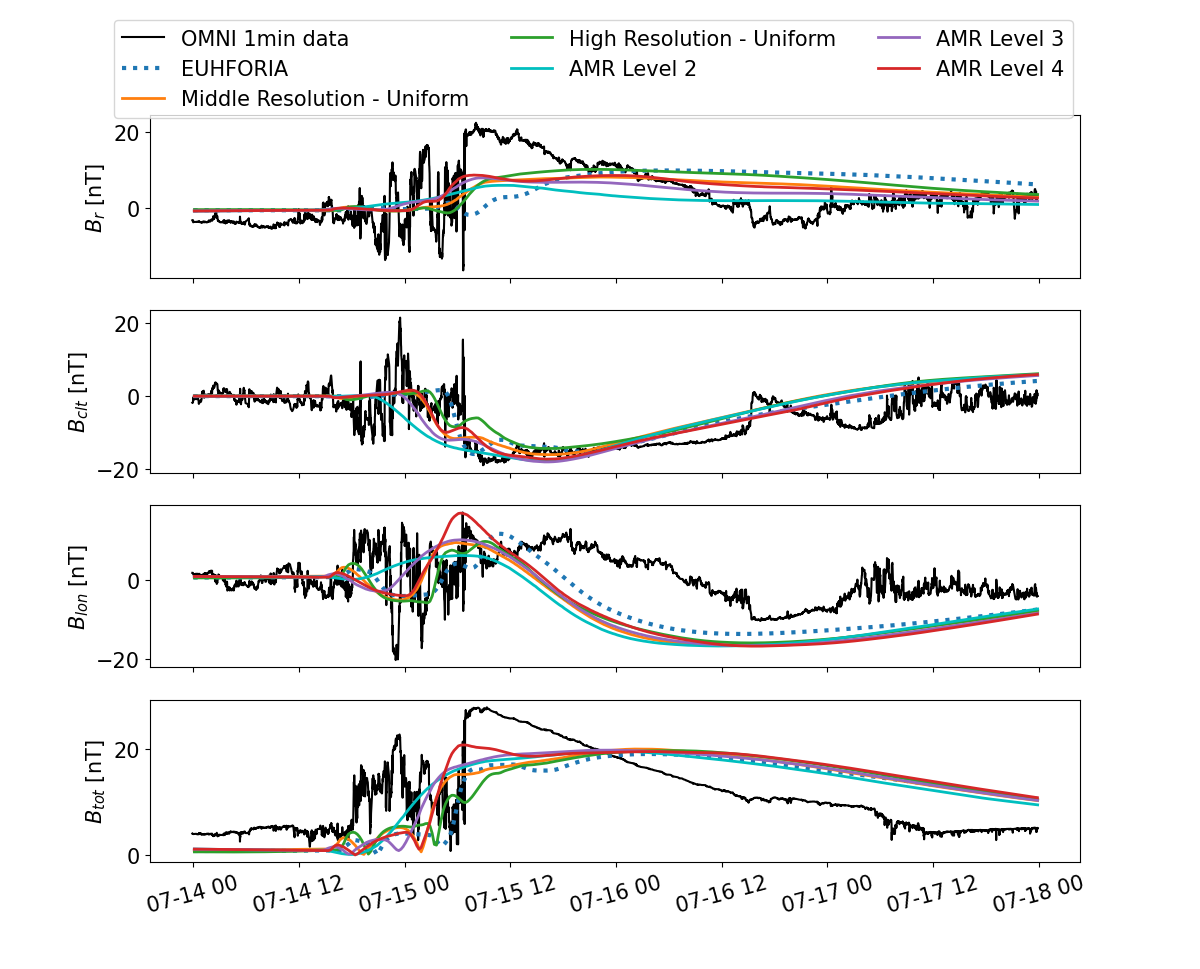}
  \caption{Time series data at Earth. Time profiles of all the magnetic field components are given for the uniform grid simulations in Icarus and EUHFORIA, together with Icarus simulations using AMR levels 2, 3, and 4. }\label{fig:B_components_timeseries}
\end{figure*}

Figure~\ref{fig:B_components_timeseries} shows the time series at Earth for the uniform grid simulations with both EUHFORIA and Icarus together with AMR level 2, 3 and 4 simulation results. All magnetic field components are plotted. Data plotted with black line corresponds to in situ OMNI-1min observational data. We noted earlier that the CME shock arrives slightly later in the EUHFORIA simulation than in Icarus simulations. This also leads to different arrival times for the magnetic clouds. The gradients in the time profiles of the plasma variables are the largest for the AMR level 4 simulation, which is showing stronger gradients than the ones obtained with the uniform grid simulation in Icarus. The arrival time is closer to the one in the OMNI data for the EUHFORIA simulation, but the strengths of the magnetic field components are slightly better modelled by Icarus AMR level 4 simulation. Overall, the profiles obtained in the Icarus simulations resemble the OMNI observed data slightly more closely, than with the EUHFORIA simulation. The simulations performed with Icarus are MHD simulations. The aim is to obtain the global picture of the evolution and arrival of the CME cloud and shock in heliosphere. Therefore, by default, our approach gives a macroscopic view. This leaves out microscopic effects like turbulence and magnetic sheath formation, but it is useful and even necessary to predict the $B_z$ component of the internal magnetic field of the CME and the shock arrival time and strength. \cite{Scolini2022_complexity} and \cite{Scolini2023} show that the internal structure of the magnetic field of flux ropes changes significantly during their propagation, and the magnetic complexity increases during this evolution. Therefore, higher resolution simulations are necessary and enable us to unravel the intricate internal plasma and magnetic structures and to model the evolution of the CMEs much more accurately. In general, the LFF-Spheromak CME model is successful to introduce the magnetized CME in the heliosphere, as the impact at Earth is mimicked relatively well, that is, the time profiles at Earth (L1) compare well to the in-situ measure profiles. More precisely, the CME shock arrival time, the strength of the magnetic field components, and thus the direction of the magnetic field, are all in line with the observational data. In the last panel of Figure~\ref{fig:B_components_timeseries} the difference between AMR simulation effects are shown clearly. The simulation result with AMR level 2 shows a smooth profile, omitting the variations in the total magnetic field strength before the actual magnetic cloud arrival. The simulation result with AMR level 3 already shows a slight increase in the total magnetic field component, similar to the observational data, and also some variation in the total magnetic field before the shock arrival. For both the AMR level 4 and uniform grid simulations in Icarus and EUHFORIA, model the magnetic cloud the most similar to OMNI data. Yet, the peak value of the total magnetic field is the largest in the AMR level 4 simulation results, coming closer to the observed value than with the other, lower resolution, simulations. We also conducted a simulation with AMR level 5, that is with 4 additional levels of refinement, in order to check how higher AMR level performs. The result is given in Appendix~\ref{appendix:timeseries}. The peak value in the AMR level 4 result is reached at the time at which the total magnetic field increases to its maximum in the observational data. In the AMR level 5 simulation results, the peak still coincides with the AMR level 4 peak, which suggests that the arrival of the magnetic cloud is modelled rather well in Icarus, and with even higher resolution simulations the shock arrival would converge to the one in the observational data. Since, the peak values in the different magnetic field components are so similar to those in the observed data, the AMR level 5 simulation is not considered necessary, AMR level 4 should be sufficient for forecasting. Notice that the considered CME is rather hard to model, since it occurred in an active phase of the Sun and the background wind has a complex structure. During its propagation, the CME interacts with a high speed stream, resulting in high gradients that need to be resolved in the computational domain. Therefore, the simulations with more AMR levels require longer wall-clock times than in less complicated cases. As it can be observed in Figure~\ref{fig:B_components_timeseries_amr5}, the gradients in the time profiles of the magnetic field components are slightly larger in the AMR level 5 results than in the AMR level 4 results, but there are no significant differences in the arrival time of the magnetic cloud nor in the strengths of the magnetic field components, to consider this additional refinement level necessary. The results, including the AMR level 5 simulation time series, are shown in figures~\ref{fig:speed_numberdensity_timeseries_amr5} and \ref{fig:B_components_timeseries_amr5} in Appendix~\ref{appendix:timeseries}.

Clearly, in the results of simulations using a higher number of AMR levels, one can distinguish more qualitative features compared to simulations results from cases with a lower number of AMR levels. In the total magnetic field strength, for instance, one can see that there is some structure being resolved in high AMR level simulation results just before the arrival of the magnetic cloud. The AMR level 4 simulation result shows stronger variations than the lower AMR level simulations. Looking at the AMR level 5 results in Fig.~\ref{fig:B_components_timeseries_amr5}, the profiles of the time evolution at L1 of the magnetic field components as well as of the total magnetic field strength, are also different from those obtained with the AMR level 4 simulation. This clearly indicates that higher AMR level simulations resolve smaller scale structures and introduce a more prominent magnetosheath in the simulation. However, elaborating more or deeper on this would require a totally different setting for the simulations, with strongly restricted AMR criteria, focused only on the area of interest in order to enable locally ultra high resolution simulations with very high AMR levels. This is rather important as the magnetosheath region is very turbulent and the oscillations in the observational data are not noise, but have a physical meaning. We consider this as a future interesting application for the Icarus heliospheric modelling tool, as standard modelling with uniform grids do not allow this kind of studies due to CPU time restrictions.
Overall, considering the chosen MHD approach, the synthetic (modelled) data with high resolution Icarus simulations meets the expectations in comparison to the OMNI observational data. As a matter of fact, the profiles for the magnetic field components have a similar evolution in time as for the OMNI data, the shock steepness increases with higher resolution simulations and the peak value is reached at a time that is comparable to OMNI data. Also in the number density profiles we can spot that, with the higher resolution simulations, the CME is more localized and less diffuse or spread out, which makes it more comparable to the OMNI data.

\begin{table}[htb!]
  \caption{Run times (wall-clock time) required for the EUHFORIA middle resolution simulation, the Icarus middle-resolution simulation with the uniform (Middle$_{EQ}$) grid, and the AMR level 2, 3, and 4 simulations. All the simulations were performed on 1 node with 2 Xeon Gold 6240 CPUs@2.6 GHz (Cascadelake), 18 cores each, on the Genius cluster at KU Leuven.}
  \centering
   \begin{tabular}{c c c c c }
  \hline\hline
   EUHFORIA & Middle$_{EQ}$ & AMR2& AMR3 & AMR4 \\[4pt]
   \hline
18h 34m & 6h 52m &  0h 11m & 0h 27m & 2h 34m \\ \hline
 \end{tabular}
  \label{table:combi_runs_times}

\end{table}

Table~\ref{table:combi_runs_times} summarizes the simulation times required for each simulation on 1 node on the Genius cluster with 2 Xeon Gold 6240 CPUs@2.6 GHz (Cascadelake), 18 cores each, at KU Leuven. The middle resolution uniform grid simulation in EUHFORIA took 18 hours and 34 minutes, while with Icarus the very similar setup takes only 6 hours and 52 minutes. The simulations with advanced numerical techniques are faster. The simulation on the stretched grid in combination with AMR level 2 takes only 11 minutes, while with AMR level 3 it takes 27 minutes and with AMR level 4 it needed 2 hours and 34 minutes. Thus, the AMR level 4 simulation was 2.67 times faster than Icarus uniform grid simulation and 7.23 times faster than the EUHFORIA uniform grid simulation.

\section{Conclusions and Outlook} \label{section:discussion_conclusion}
A magnetized CME model represented by a Linear Force-Free Spheromak has been implemented in the new heliospheric wind and CME evolution model Icarus and is validated in this study. The effect of advanced numerical techniques such as gradual radial grid stretching and solution adaptive mesh refinement are considered. 

Icarus is a 3D MHD heliospheric wind and CME evolution model, covering the same domain as EUHFORIA, solving the ideal MHD equations on uniform spherical or optimized grids with a finite volume numerical scheme. The coordinate frame in Icarus is co-rotating with the Sun, while in EUHFORIA the equations are solved in the HEEQ coordinate system (in which the longitude of the Earth is fixed, but its latitude and radial distance vary). The model takes the output of the WSA-like coronal model in EUHFORIA and uses it as inner boundary conditions to determine the heliospheric wind. After an MHD relaxation phase, a stationary solar wind is obtained in the simulation, on which CMEs can be superposed by injecting them at the inner boundary, at 0.1~AU. 

In this study, the implementation of a magnetized CME model is examined. The LFF-Spheromak model is imported in Icarus from EUHFORIA via coupling between the two architectures. The diagram given in Figure~\ref{fig:diagram_linking} shows how the linking is set up between the C++ and Fortran codes. This is achieved by generating a library from the C++ code, calculating the internal magnetic field components of the spheromak CME, which can be accessed from the Fortran code of Icarus. This way, the injected CME in Icarus is guaranteed to be identical to the one in EUHFORIA and, moreover, additional repetitive work is avoided. The LFF-Spheromak CME model in EUHFORIA is represented by the definition for the magnetic field that is divergence-free and force-free (unlike the Gibson and Low spheromak model). The equations describing such configuration are only briefly discussed in Section~\ref{section:spheromak_model}, since the original model was implemented in EUHFORIA by \cite{Verbeke2019}. 

Advanced numerical techniques are applied to optimize the computational grid in Icarus. The standard uniform spherical grid is of course available, similar to the one in EUHFORIA. Additionally, gradual radial grid stretching and AMR can be applied. The radial grid stretching preserves the aspect ratio between the width and the length of the cells. Therefore, the cells are not as deformed as in the uniform spherical grid simulations, near the inner and outer boundaries. On the other hand, the block-adaptive AMR guarantees high spatial resolution in the domain only in locations where this is necessary. The refinement condition(s) is/are controlled by the user, leading to major flexibility in optimizing the computational grid on which the simulation is performed. Also, the maximum number of refinement levels can be specified. Whenever the implemented AMR criteria are met, the designated block in the computational grid is refined to the prescribed maximum refinement level. Moreover, when the conditions are not met any more, the grid is coarsened again, to avoid unnecessary CPU use.

In this study, the magnetized LFF Spheromak CME was injected in Icarus simulations using different grids. The goal of the work is to focus on the macroscopic effects, such as the global structure of the CME, its shape, its complex internal magnetic field structure,  and its associated shock. The results were compared to the similar simulation with EUHFORIA, using the same magnetogram and inner boundary conditions and with the same settings. Apart from the original uniform grid, several combinations of radial grid stretching and different AMR levels were considered. A real event study was performed in order to validate the model. The CME of July 12, 2012 was modelled with the pre-determined parameters available in the literature. The appropriate GONG magnetogram was chosen as inner boundary condition for the empirical WSA-like coronal model, to provide the plasma conditions at the inner boundary of the heliospheric model (at 0.1~AU). The AMR criteria were the same in all the AMR simulations and consisted of a combination of two refinement criteria, which yield high spatial resolution in both the CME interior and the CME shock. The refinement was also limited to the neighbourhood of the Sun-Earth line for optimization of forecasting simulations. The results were analysed using the time series data of the MHD quantities at Earth and 1~AU slice snapshots. Clearly, the surrounding solar wind is best resolved in uniform grid simulations both with Icarus and EUHFORIA, since this grid is uniform and the resolution is relatively high everywhere in the domain, independently, whether it is a CME region or the solar wind. In simulations using solution AMR, the grid is adjusted depending on the solution and the specified criteria. Since we are focusing on the CME, the resolution is only increased in and around the CME and the surrounding solar wind displays smoother profiles. On the other hand, the CME interior is far better resolved in the AMR simulations. With increasing the maximum number of AMR levels, the observed structures inside the CME are finer and the boundary between inversely polarized regions is narrower. More details can be seen in the AMR level 4 simulation result, compared to the AMR level 2, 3 and uniform grid simulation results. The time series profiles at Earth, as the CME passes through the position of Earth, give a good estimation of how well resolved different regions in the CME are at 1~AU for the different simulations. For assessing the performance of the combined AMR criterion, the radial velocity component and number density profiles are shown in Figure~\ref{fig:speed_numberdensity_timeseries} and the magnetic field components in Figure~\ref{fig:B_components_timeseries}. The shock arrival is best resolved by the AMR level 4 simulation, since the AMR criterion includes the shock regions in the domain. The profiles are steeper than in the other simulations, converging to the observed data steepness profile with higher numbers of AMR levels. The magnetic field components of the spheromak CME are also best resolved in with AMR level 4, since the combined criterion was also aimed at resolving the CME interior. The strength and profiles of the magnetic field components modelled by Icarus are slightly more similar to the OMNI data than the data modelled by EUHFORIA. The magnetic cloud arrives somewhat later in the EUHFORIA simulation, which is closer to the arrival data in OMNI data, but the strength of the magnetic field in the magnetic cloud is better estimated by the Icarus model. 

In conclusion, the simulation results are very similar to those obtained with EUHFORIA, but require much less CPU time. The arrival time was slightly better modelled by EUHFORIA, but the profiles of the different variables considered (radial velocity, number density and magnetic field components) are slightly better modelled in Icarus simulation. The AMR level 4 simulation (with 3 refinement levels) produced the best results and when comparing to the AMR level 5 simulation results, with 4 refinement levels, it became clear that AMR level 4 is sufficient for modelling the spheromak CMEs with the purpose to accurately estimating the arrival time of the shock and the magnetic field component configurations. AMR level 5 produces slightly sharper shock results, but not significantly. Therefore, it is not considered worthwhile, in consideration of the longer computational times required. 

Table~\ref{table:combi_runs_times} summarizes the wall-clock time each simulation required on 1 node of Genius cluster at KU Leuven. The uniform grid simulations with EUHFORIA and Icarus, having relatively high spatial resolution everywhere in the domain, need $\sim$ 18.5 and $\sim$ 7 hours, respectively. The simulations combining radial grid stretching and AMR levels 2, 3 and 4, require $\sim$ 0.2, 0.45 and $\sim$ 2.5 hours, respectively. The AMR level 4 simulation showed the best results, while being $\sim 2.7\;$times faster than the uniform grid simulation in Icarus and 7 times faster than the EUHFORIA simulation. The speed-up obtained by the comparison of these simulations is the modest, because the most broad refinement criterion is used, which refines both the shock and the CME interior. Simulations only focusing on the CME shock are much faster, since the CME interior area is not taken into consideration for high resolution in that case. The AMR level 3 model data is the most similar to the uniform grid, standard resolution simulations in Icarus and EUHFORIA. The AMR level 3 simulation is $\sim$ 15 and $\sim$ 41 times faster than the  uniform grid with middle resolution simulations in Icarus and EUHFORIA, respectively. The timings of the simulations are not constant, other real event simulations with different magnetic field configuration of the CME and different solar wind, might require different wall-clock times for computations. However, the case considered here is close to the solar maximum, leading to quite variable solar wind and the modelled CME has high speed, causing strong shocks along its propagation, and in addition it interacts with a co-rotating interaction region. Therefore, this event study is a good estimation of how the model performs in complex cases. 

The newly introduced magnetized CME model in Icarus enables in-depth studies of the effects of the interactions of the CMEs with the magnetic field of the background solar wind, e.g.\ deformation, deflection and flux and plasma erosion. With the previous simple cone CME model, only the HD variables could be examined. Therefore, only the arrival time and the shock strength at Earth could be estimated. Since spheromak model includes the internal magnetic field configuration, the geo-effectiveness of the CME interaction with the Earth magnetic field can be studied, e.g.\ by calculating geomagnetic indices. The evolution of the complexity of the internal magnetic field of the CME can also be examined all the way from 0.1~AU to Earth and beyond.  A multi-spacecraft study is considered with the magnetized CME model in the future, as it will enable to validate the model at multiple points and the magnetic field can be compared at different heliospheric locations. The combination of advanced numerical techniques with the magnetized CME model provides an opportunity to perform multiple simulations and investigate the effect of different parameters, without spending too many computational resources. This leads to a deeper insight in the evolution and propagation of the magnetized CMEs all the way to Earth, which, on the other hand, will help to forecast such events more accurately and mitigate the possible damage more effectively.

\begin{acknowledgements}
This research has received funding from the European Union’s Horizon 2020 research and innovation programme under grant agreement No 870405 (EUHFORIA 2.0) and the ESA project "Heliospheric modelling techniques“ (Contract No. 4000133080/20/NL/CRS).
These results were also obtained in the framework of the projects C14/19/089  (C1 project Internal Funds KU Leuven), G.0B58.23N and G.0025.23N  (FWO-Vlaanderen), SIDC Data Exploitation (ESA Prodex-12), and Belspo project B2/191/P1/SWiM.
The Computational resources and services used in this work were provided by the VSC-Flemish Supercomputer Center, funded by the Research Foundation Flanders (FWO) and the Flemish Government-Department EWI.
\end{acknowledgements}

%
%

\bibliographystyle{aa}
\bibliography{bibliography}
\begin{appendix} 
\section{Time series in Icarus} \label{appendix:timeseries}
\subsection{Time series with AMR level 5}

\begin{figure}[hbt!]
    \includegraphics[width=\linewidth]{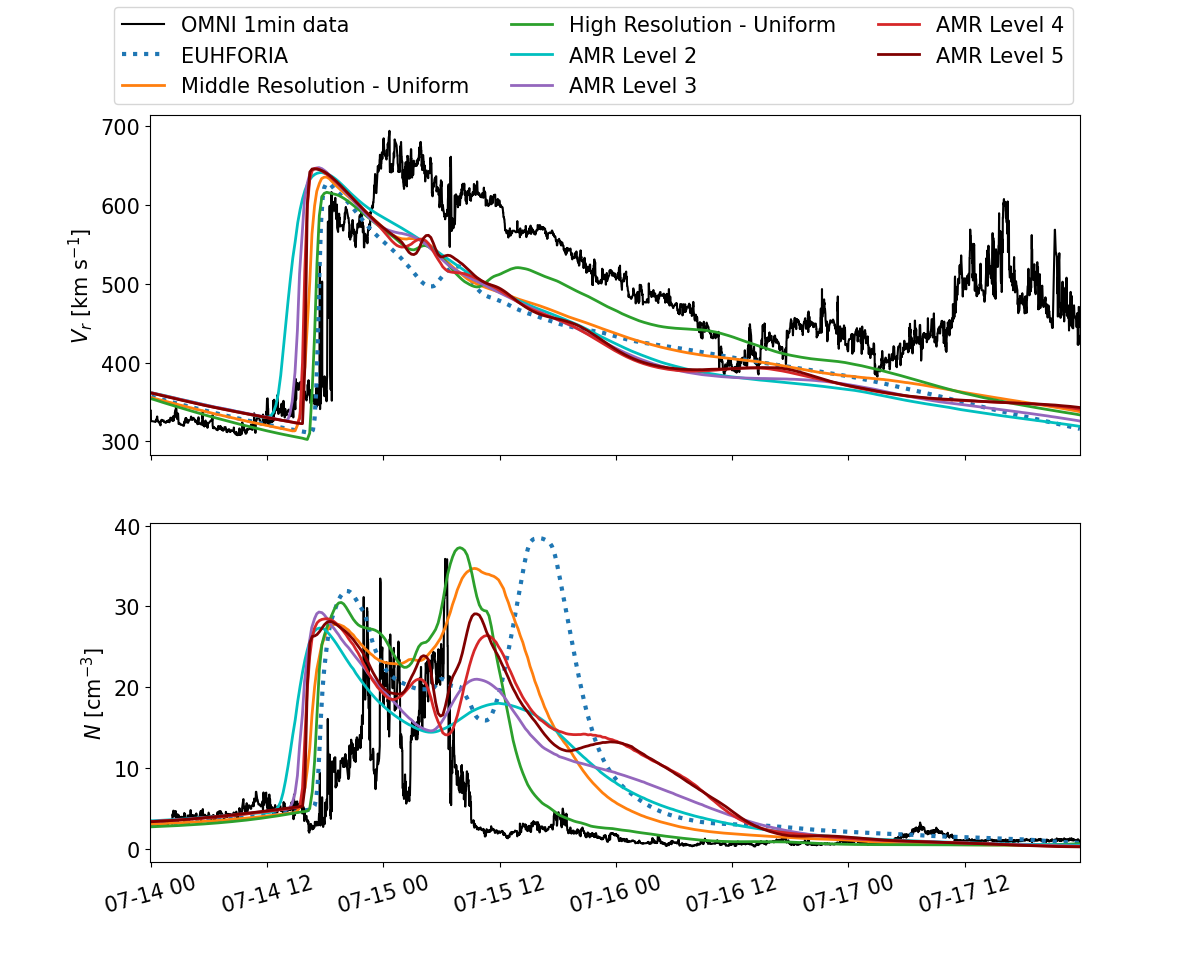}
    \caption{Time series data extracted at the location of Earth. Radial speed and number density values are given for the uniform grid simulations in Icarus and EUHFORIA, together with Icarus simulations using AMR levels 2, 3, 4 and 5.} \label{fig:speed_numberdensity_timeseries_amr5}
\end{figure}

\begin{figure*}[hbt!]
\centering
    \includegraphics[width=\linewidth]{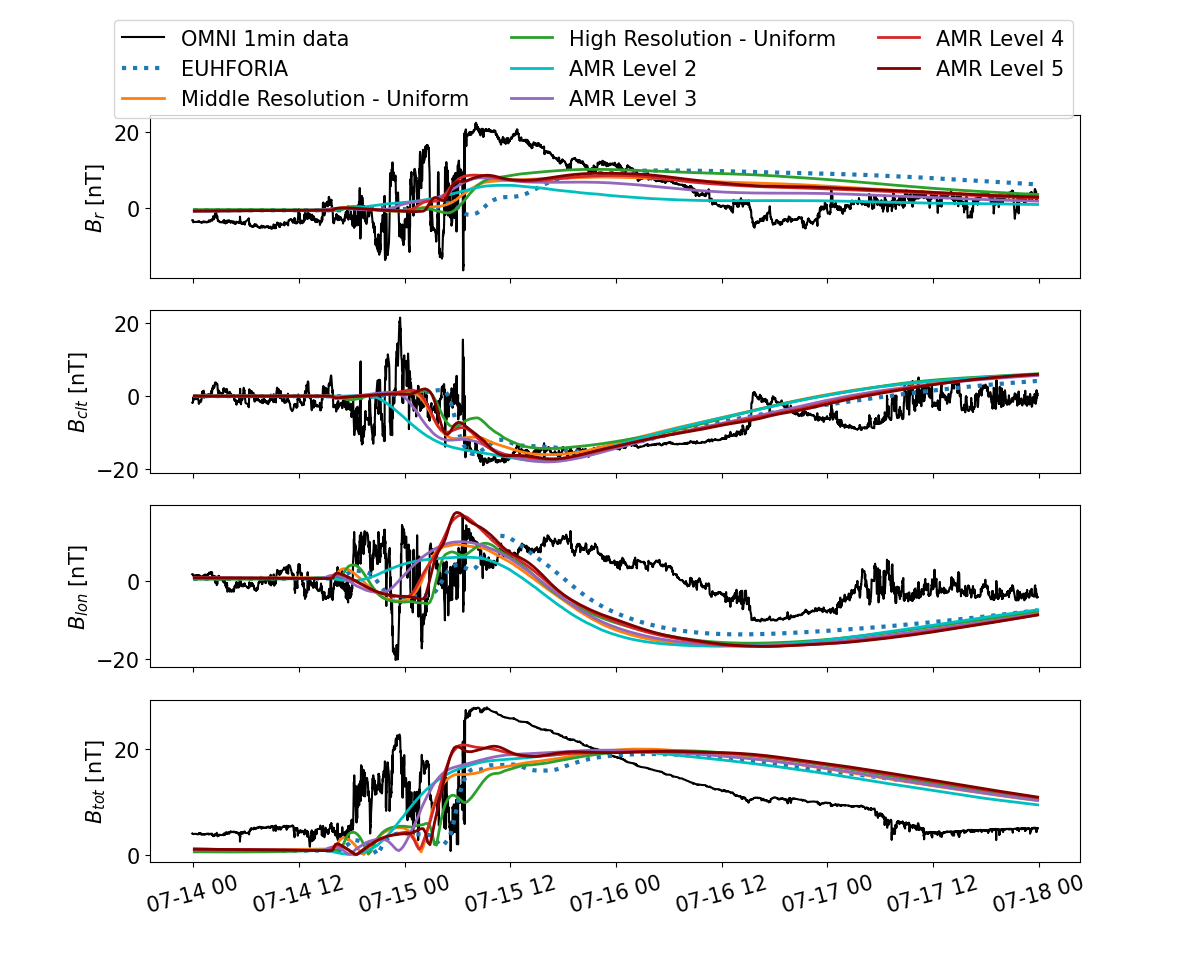}
  \caption{Time series data at Earth. All magnetic field components are given for the uniform grid simulations in Icarus and EUHFORIA, together with Icarus simulations using AMR levels 2, 3, 4 and 5. }\label{fig:B_components_timeseries_amr5}
\end{figure*}

Figures~\ref{fig:speed_numberdensity_timeseries_amr5} and \ref{fig:B_components_timeseries_amr5} show time series profiles at Earth, including the AMR level 5 simulations results. The colours denote the same simulations as in Figure~\ref{fig:speed_numberdensity_timeseries}, where here, additionally, the AMR level 5 simulation result is plotted in purple. The AMR level 5 curve is almost identical to the AMR level 4 result, especially regarding the shock arrival time and the peak speed upon arrival. It displays more features as the CME is passing through Earth in the number density values. The features are also present in the AMR level 4 result, but less pronounced. 

The magnetic field components also show more features with highly resolved grids. The AMR level 5 simulation result displays sharper profiles in the magnetic field temporal variations, visible in all the magnetic field components. Although, the magnetic field strength time profile is the same as that modelled with the AMR level 4 simulation. The profiles are also very similar most of the part, the difference is mostly present upon the arrival of the magnetic cloud. In the time profile of the $B_\text{r}$-component, there is a more pronounced bump in the AMR level 5 curve before the large increase in the magnitude upon the magnetic cloud arrival. It is also visible in the AMR level 4 curve, but the profile is more smoothed out. In the time profile of the $B_\text{clt}$ component, the AMR 4 and 5 curves are rather similar, with slightly higher gradients present upon the magnetic cloud arrival in the AMR level 5 result. In the time profile of the B$_\text{lon}$ component, the maximum value of the component is modelled well only by the AMR level 4 and 5 simulations. Similarly, in the time profile of the total magnetic field strength, the strongest magnetic field is obtained in the simulations with the most AMR levels, stronger than in all other simulations, including the uniform grid Icarus and EUHFORIA simulations. 

\end{appendix}
\end{document}